%Paper: hep-th/9401044
%From: Josh Boorstein <boo7@bohr.uchicago.edu>
%Date: Tue, 11 Jan 94 13:28:12 CST

\input harvmac

\def\np{Nucl. Phys. }

\def\pr{Phys. Rev. }

\def\ch{{\cal H}}

\def \cl{{\cal L}}
\def \cm{{\cal M}}
\Title{\vbox{\hbox{EFI-93-69}
\hbox{\tt hep-th/9401044}}}
{{\vbox {\centerline{Symmetries and Mass Splittings}
\bigskip
\centerline{in QCD$_2$ Coupled to Adjoint Fermions}
}}}

\centerline{\it Joshua Boorstein and David Kutasov}
\smallskip\centerline
{Enrico Fermi Institute}
\centerline {and Department of Physics}
\centerline{University of Chicago}
\centerline{Chicago, IL 60637, USA}
\vskip .2in

\noindent
Two dimensional QCD coupled to fermions in the adjoint
representation of the gauge group $SU(N)$, a useful toy model
of QCD strings,  is supersymmetric
for a certain ratio of quark mass and gauge coupling constant.
Here we study the theory in the vicinity of the supersymmetric
point; in particular we exhibit the algebraic structure
of the model
and show that the mass splittings as one moves away from the
supersymmetric point obey a universal relation of the form
${M_i}^2(B)-{M_i}^2(F)=M_i\delta m+O(\delta m^3)$. We discuss
the connection of this relation to string and quark model expectations and
verify it numerically for large $N$. At least for low lying
states the $O(\delta m^3)$ corrections are extremely small.
We also discuss a natural generalization of QCD$_2$ with an infinite
number of couplings, which preserves SUSY. This leads to a Landau --
Ginzburg description of the theory, and may be useful for defining
a scaling limit in which smooth worldsheets appear.

\Date{12/93}
%\draftmode
%

\newsec{Introduction.}

Two dimensional Yang -- Mills theory coupled to adjoint matter
has been argued to be  an interesting toy model for studying
QCD strings
\ref\kl{S. Dalley and I. Klebanov, \pr {\bf D47} (1993) 2517.},
\ref\ku{D. Kutasov, Chicago Preprint
EFI-93-30, hep-th/9306013, Nucl. Phys
{\bf B}, in press.},
being probably the simplest confining gauge theory which
undergoes a deconfining transition in the leading order
in the $1/N$ expansion \ku.
In string theory there is a qualitative difference
between models with an exponential growth in the density
of states with mass (and consequently a Hagedorn transition), and
ones where such growth is absent
\ref\ks{D. Kutasov and N. Seiberg, \np {\bf B358} (1991) 600.}
-- the two are separated by
the famous $c=1$ barrier.
It is natural to expect a similar ``transition'' in gauge theory as well.
Since large $N$ gauge theories with finite densities of states
are known to be described by strings
\ref\gm{P. Ginsparg and G. Moore,
1992 TASI lectures, hep-th/9304011.},
\ref\gross{D. Gross, Nucl. Phys. {\bf B400} (1993) 161;
D. Gross and W. Taylor, Nucl. Phys. {\bf B400} (1993) 181.},
it seems important to ``cross the $c=1$ barrier'' in gauge theory and
understand
the nature of the relation to strings in that regime.
QCD$_2$ coupled to adjoint matter provides a (hopefully)
simple case in which this regime can be quantitatively studied.

In a recent paper \ku\ it has been shown that some aspects of the model
hint at a stringy structure. In particular, the masses of certain winding modes
around compact (Euclidean) time exhibit an interesting dependence on
the parity of the winding number,
which is difficult to understand in gauge theory but is very natural
in string theory; the model with adjoint fermions exhibits an
in general  softly
broken supersymmetry (SUSY), reminiscent of similar
results in string theory \ks; and the spectrum
of the appropriate $2d$ Coulomb potential (ignoring pair production)
contains an infinite number of ``Regge trajectories''
with an exponential density
of (bosonic and fermionic) bound states, at high mass.

In this paper we are going to focus on the supersymmetry found in \ku\ for
QCD$_2$ coupled to adjoint fermions, and its (explicit) breaking by the
quark mass term.
The main questions we wish to address here are the algebraic structure
of the supersymmetric theory and possible generalizations thereof,
with the hope that the supersymmetric theory may be
exactly solvable (perhaps at large $N$), and the structure of the
theory near the supersymmetric point.  This should shed light
on questions like the applicability of stringy constraints on the spectrum
found
in \ks\ to (this) gauge theory.

In Section 2 we describe QCD coupled to quarks
in the adjoint representation of $SU(N)$,  the light -- cone
quantization of the model, and the SUSY which arises
for a specific ratio of the quark mass and the gauge
coupling.
In Section 3 we examine the theory perturbed
away from the supersymmetric point by a change of the mass of the constituent
quarks.  We
show that
the splitting in the mass squared of the super -- partner bound states is
universal, at least to second order in the deviation of the quark
mass from the supersymmetric mass.
This allows us to examine the
behavior of:
\eqn\zb{Z(\beta)=Tr (-)^{F} e^{-\beta M^2}}
near the supersymmetric point,
which plays an important role in string theory
\ks.
Section 4 contains some numerical results; we verify the universal
mass splitting by numerically calculating the mass of the lowest
lying excitations in the bosonic and fermionic sectors
as a function of the constituent quark mass, and studying the appropriate
differences. In addition we study numerically the distribution of certain signs
needed for the evaluation of \zb.
In section 5 we embed QCD$_2$ in an infinite dimensional space
of theories all of which are supersymmetric.
These theories are parametrized by a superpotential $W(\Phi)$
and exhibit an intriguing connection to (super --) Landau -- Ginzburg
theories, a fact that may be useful to understand them better.
We conclude in section 6.

\newsec{QCD$_2$ Coupled to Adjoint Fermions.}

Consider
the theory of real (Majorana) fermions in the adjoint representation of
the gauge group $SU(N)$, described by the Lagrangian,
\eqn\a{\cl={1\over 8g^2} F_{\mu\nu}^2+\bar \psi\gamma^\mu D_\mu
\psi+m\bar\psi\psi}
where $F_{\mu\nu}=\partial_{[\mu}A_{\nu]}+[A_\mu, A_\nu]$,
$D_\mu\psi=\partial_\mu\psi+i[A_\mu, \psi]$, $\psi_{ab}$ is a traceless
hermitian anticommuting matrix, $m$ is the (bare) fermion mass and $g$
the gauge coupling.
We follow the conventions of \ku\
which we will review next to establish the
notation.

Two dimensional gauge theories look especially
simple in light -- cone quantization
\ref\th{G. 't Hooft, Nucl. Phys. {\bf B72} (1974) 461.},
\ref\lc{
H. Pauli and S. Brodsky, \pr {\bf D32} (1985) 1993, 2001;
K. Hornbostel, S. Brodsky and H. Pauli, \pr {\bf D41} (1990) 3814.},
which has been recently applied to this model in
\kl, \ku.
We denote by
$\psi_{ab}$ the right moving fermions and by $\bar\psi_{ab}$ the left moving
ones.
The $SU(N)$ currents, $J^+_{ab}=\psi_{ac}\psi_{cb}$, $J^-_{ab}=
\bar\psi_{ac}\bar\psi_{cb}$ form right and left moving
level $N$ affine Lie algebras, respectively. In the
gauge $A_-^{ab}=0$ the Lagrangian \a\ takes the form:
\eqn\u{\cl={1\over 4g^2}(\partial_-A_+)^2+i\psi\partial_+
\psi+i\bar\psi\partial_-
\bar\psi-2im\bar\psi\psi+A_+J^+.}
The equations of motion for $A_+$, $\bar\psi$ do not involve
derivatives with respect to $x^+$, the
light -- cone ``time"; it is easy to integrate them out to obtain
an action solely in terms of the right moving fermions $\psi$:
\eqn\v{\cl_\psi=i\psi\partial_+\psi+g^2J^+{1\over\partial_-^2}J^++
im^2\psi{1\over\partial_-}\psi.}
Quantization on constant $x^+$ surfaces gives rise to the
momentum operator:
\eqn\w{\eqalign{P^+=&i\int dx^-\psi\partial_-\psi\cr
                P^-=&\int dx^-\left(-im^2\psi{1\over\partial_-}\psi-g^2
J^+{1\over\partial_-^2}J^+\right).\cr}}
Expanding $\psi(x^+=0)$ in modes:
\eqn\x{\psi_{ab}(x^-)={1\over2\sqrt\pi}\int_{-\infty}^\infty
dk\psi_{ab}(k) e^{-ikx^-}}
and imposing the canonical anticommutation relation,
$\left\{\psi_{ab}(x^-), \psi_{cd}(0)\right\}
=\half\delta(x^-)\delta_{a,d}\delta_{c,b}$
we find the mode anticommutation relations:
\eqn\z{\left\{\psi_{ab}(k), \psi_{cd}(k^\prime)\right\}
=\delta(k+k^\prime)\delta_{a,d}\delta_{c,b}.}
$\psi_{ab}(k)$ with $k\le0$ are creation operators, whereas the ones with
$k\ge0$ are annihilation operators. The light -- cone vacuum is chosen
such that:
\eqn\aa{\psi_{ab}(k)|0\rangle=0\;\;\;\;\;\forall k>0.}
The momentum operators \w\ are normal ordered in the standard fashion,
and take the form:
\eqn\ab{\eqalign{P^+=&\int_0^\infty dk k\psi_{ab}(-k)\psi_{ba}(k)\cr
                 P^-=&m^2\int_0^\infty{dk\over k}\psi_{ab}(-k)\psi_{ba}
(k)+g^2\int_0^\infty{dk\over k^2} J^+_{ab}(-k)J^+_{ba}(k)\cr}}
where $J_{ab}^+(k)$ is given by (for $k\not=0$):
\eqn\ac{J^+_{ab}(k)=
\int_{-\infty}^\infty dp \psi_{ac}(p)\psi_{cb}(k-p).}
$J_{ab}^+(0)=\int_0^\infty dp\psi_{ac}(-p)\psi_{cb}(p)$
must annihilate
all physical states (due to confinement).
Physical states
are obtained by acting with raising operators $\psi_{ab}(-k)$ on the vacuum
\aa. $P^+$ is diagonal in this basis; therefore, eigenmodes
of $P^-$ are also eigenmodes of the mass operator, $M^2\equiv 2P^+P^-$.
Since $P^+$ commutes with all the
operators mentioned below, we will usually set
it to $1$ from now on to simplify some formulae.

It was pointed out in \ku\ that the dynamics described by \z, \ab\ is
supersymmetric, at least for $m^2=g^2N$.
In particular, the operator:
\eqn\ad{Q^+={1\over{3\sqrt{N}}}\int dx^-\psi_{ab}\psi_{bc}\psi_{ca}
={1\over{3\sqrt{N}}}\int_{-\infty}^\infty dp
\psi_{ab}(-p)J_{ba}^+(p)}
was shown in \ku\ to commute with the light -- cone Hamiltonian:
\eqn\af{[Q^+, P^-]=0,\;\;\;{\rm for}\;\; m^2=g^2N. }
In addition, a simple calculation shows that:
\eqn\ag{(Q^+)^2=P^+.}
Thus $Q^+$ is a conserved charge.
The supersymmetry transformation:
\eqn\ae{\eqalign{\delta\psi_{ab}(p)=& \epsilon\{Q^+,\psi_{ab}(p)\}={\epsilon
\over\sqrt{N}} J_{ab}^+(p)\cr
\delta J_{ab}^+(p)=&
\epsilon[Q^+,J_{ab}^+(p)]=-\epsilon\sqrt{N}p\psi_{ab}(p)\cr}}
acts non-linearly on $\psi$.

Clearly \ad\ can not be the full symmetry
of the light -- cone Hamiltonian. The theory we are discussing
is left -- right symmetric, and although the chiral gauge chosen ($A_-=0$)
makes the symmetry non -- manifest, physics must
be symmetric; hence, there must exist another supercharge
$Q^-$ with the properties (for $m^2=g^2N$):
\eqn\aff{[Q^-, P^\pm]=0,\;\;\;
(Q^-)^2=-P^-}
(the minus sign in the last of equations \aff\ will be convenient later.)
Indeed, we shall show shortly that the appropriate conserved charge is:
\eqn\qm{\eqalign{Q^{-}=&-{g\over3}\int{dp_1dp_2dp_3}\delta(p_1+p_2+p_3)
({1\over{p_1}}+{1\over{p_2}}+{1\over{p_3}})\psi_{ab}(p_1)\psi_{bc}(p_2)
\psi_{ca}(p_3)\cr
=&g\int_{-\infty}^\infty {dp\over{p}}
\psi_{ab}(-p)J_{ba}^+(p).\cr}}

One can of course verify that $Q^-$ \qm\ indeed satisfies \aff\
by explicit calculation, but that involves careful normal ordering
and is somewhat tedious. A simpler algebraic derivation which is
also useful for the next sections is the following.
Consider the light -- cone Hamiltonian of the theory $P^-$ \ab, written
in the form:
\eqn\sss{P^-=P^-_{\rm susy}+\alpha H_0}
where $P^-_{\rm susy}$ is the supersymmetric Hamiltonian ( \ab\ with
$m^2=g^2N$), $\alpha=m^2-g^2N$, and:
\eqn\ho{H_0 = \int_0^\infty{dk\over k}\psi_{ab}(-k)\psi_{ba}(k).}
Now define an operator $F(\gamma)$ by
\eqn\fa{F(\gamma) = \exp(\gamma H_0) Q^+ \exp(-\gamma H_0).}
Expanding $F$ in powers of $\gamma$ we have:
\eqn\fx{F(\gamma) = \sum_{n=0}^\infty {{\gamma^n}\over{n!}} Q_n}
where $Q_n$ is given by the standard Baker-Hausdorff formula, i.e.
\eqn\bhf{Q_n = [H_0,[H_0,[H_0,...[H_0,Q^+]]]...].}
Writing out $Q_n$ explicitly we have
\eqn\qn{Q_n={{(-)^n}\over{3\sqrt{N}}}\int{dp_1dp_2dp_3}{\delta(p_1+p_2+p_3)}
{({1\over{p_1}}+{1\over{p_2}}+{1\over{p_3}})^n}{\psi_{ab}(p_1)\psi_{bc}(p_2)
\psi_{ca}(p_3)}}
where the integration extends over the complete three dimensional space.
Note that:
\eqn\qone{Q^-=g\sqrt{N}Q_1.}

Now we note that $F^2(\gamma) = P^+$ is independent of $\gamma$.
Requiring that $F^2(\gamma)$ \fx\ should not depend on $\gamma$
gives rise to an
infinite tower of anticommutation relations satisfied by the $Q_n$'s.
The order $\gamma$
relation is $\{Q^+,Q_1\}=\{Q^+, Q^-\}=0$. At order $\gamma^2$ we find:
\eqn\qzqt{\{Q^+,Q_2\}=-\{Q_1,Q_1\}.}
After some elementary algebra
it is possible to write $Q_2$ in the form:
\eqn\qtw{Q_2={1\over{\sqrt{N}}}\int_{-\infty}^\infty{dp\over{p^2}}
\psi_{ab}(-p)J_{ba}^+(p).}
Anti-commuting this operator with $Q_0(=Q^+)$ with the help of
\ae\ and using the
derived relationship $\qzqt$ we find that
$(Q^-)^2=-P^-$ as expected.

To summarize, we find that QCD$_2$ coupled to adjoint matter
at its supersymmetric point, $m^2=g^2N$, has a standard $(1,1)$ SUSY
algebra,
\eqn\sus{(Q^\pm)^2=\pm P^\pm;\;\;\;\{Q^+,Q^-\}=0.}
Note the hermiticity properties of the supercharges \ad, \qm:
$(Q^\pm)^\dagger=\pm Q^\pm$. This is the origin of the peculiar signs
in eqs \aff, \sus.

\newsec{The vicinity of the supersymmetric point.}

The light -- cone Hamiltonian $P^-$ \ab\ can be parametrized
as:
\eqn\ssss{P^-(\alpha)=-(Q^-)^2+\alpha H_0}
with $\alpha$ as in \sss\ the deviation of the constituent
quark mass squared from its supersymmetric value $g^2N$,
and $H_0$ given by \ho. At $\alpha=0$ the spectrum is supersymmetric;
bosonic states $|B\rangle$ are paired with fermionic ones,
$|F\rangle=Q^+|B\rangle$:
\eqn\specsus{P^-_{\rm susy}|B\rangle=M^2|B\rangle;\;\;\;
P^-_{\rm susy}Q^+|B\rangle=M^2Q^+|B\rangle.}
Actually, one can do slightly better and diagonalize the ``mass''
operator $\cm\equiv Q^+ Q^-$, which satisfies
$\cm^2=P^-_{\rm susy}$. Eigenstates of $\cm$ satisfy:
\eqn\qpm{Q^-|B\rangle=MQ^+|B\rangle.}
Here $M$ can be either positive or negative, and the relative signs
of the $M$'s corresponding
to different bound states $|B\rangle$ will actually
play a role later.

At any rate, as we turn on $\alpha$ in \ssss, the masses of the
degenerate bosons and fermions \specsus\ are expected to change in a
complicated way. However, as we shall now show, the mass {\it splittings}
exhibit a simple universal behavior, at least to second
order in
$\alpha$. Indeed, consider a bosonic eigenstate of the light --
cone Hamiltonian $P^-(\alpha)$:
\eqn\bbb{P^-(\alpha)|B\rangle_\alpha=M_B^2(\alpha)|
B\rangle_\alpha.}
In general, there is no reason for $|F\rangle_\alpha\equiv
Q^+|B\rangle_\alpha$ to be an eigenstate of $P^-(\alpha)$
(of course, this is the case at $\alpha=0$ due to supersymmetry
\specsus). Nevertheless, it turns out that $|F\rangle_\alpha$
is actually an eigenstate of $P^-(\alpha)$ {\it to first order
in $\alpha$}. To verify that, we compute (unless stated otherwise,
we put $g\sqrt{N}=1$ from now on):
\eqn\comm{\left[P^-(\alpha)-M_B^2(\alpha)\right]|F\rangle_\alpha=
\left[P^-(\alpha), Q^+\right]|B\rangle_\alpha=\alpha
\left[H_0, Q^+\right]|B\rangle_\alpha=
\alpha Q^-|B\rangle_\alpha}
where in the last step we have used a result from section 2:
\eqn\fund{Q^-=\left[H_0, Q^+\right].}
Now, to first order in $\alpha$ we can replace $\alpha Q^-|B\rangle_\alpha
\rightarrow\alpha Q^-|B\rangle_{\alpha=0}$ or, by \qpm:
$$\alpha Q^-|B\rangle_\alpha=\alpha M_BQ^+|B\rangle_\alpha+O(\alpha^2)=
\alpha M_B|F\rangle_\alpha+O(\alpha^2).$$
Substituting this in \comm, we find that $|F\rangle_\alpha$ is indeed
an eigenstate to this order, and the mass splitting is:
\eqn\ssbc{M_F^2(\alpha)-M_B^2(\alpha)=\alpha M_B +O(\alpha^2).}
This can also be written in the form:
$M_F(\alpha)-M_B(\alpha)={\alpha\over{2}}+O(\alpha^2)$,
where we remind the reader that the masses
$M$ can be positive or negative depending
on the sign in \qpm. Thus, at least to first order in $\alpha$ the mass
splittings in this model are highly universal.

What happens at higher orders in $\alpha$? One can easily show using standard
techniques that if $|B\rangle_\alpha$ and $|F\rangle_\alpha$ are eigenstates
of $P^-(\alpha)$, \ssss\ with eigenvalues $M_B^2(\alpha)$,
$M_F^2(\alpha)$ respectively, then
\eqn\allord{M_F^2(\alpha)-M_B^2(\alpha)=\alpha{{}_\alpha\langle F|
Q^-|B\rangle_\alpha\over
{}_\alpha\langle F|Q^+|B\rangle_\alpha}}
To find the mass splitting to second order in $\alpha$ we need
to keep terms up to first order in the ratio of inner products
in \allord. To that order, we can substitute $|F\rangle_\alpha=
Q^+|B\rangle_\alpha$ (see \comm, \fund), so:
\eqn\alld{M_F^2(\alpha)-M_B^2(\alpha)=\alpha{{}_\alpha\langle B|
Q^+Q^-|B\rangle_\alpha\over
{}_\alpha\langle B|B\rangle_\alpha}+O(\alpha^3)}
Using \qpm\ it is easy to see that \alld\ implies that
the order $\alpha^2$ term in $\delta M^2$ vanishes and, finally,
\eqn\ssbd{M_F^2(\alpha)-M_B^2(\alpha)=\alpha M_B +O(\alpha^3).}

As mentioned in the introduction, and explained in \ku,
it would be very interesting
to calculate the partition sum $Z(\beta)$ \zb\ (at least at large $N$).
In conventional string theory this would have the property that
$\lim_{\beta\to 0}Z(\beta)=$ finite, despite the fact that the separate
contributions of bosons and fermions to \zb\ diverge as $Z_B(\beta),
Z_F(\beta)\simeq \beta^a\exp(b/\beta)$ as $\beta\to 0$.
It is far from clear in QCD that $Z(\beta\to 0)$ is indeed
(or should be) finite. To first order in $\alpha$, the calculation
of this quantity reduces to:
\eqn\zalp{Z(\beta)={\rm Tr}_{B,F}(-)^Fe^{-\beta M^2(\alpha)}=-
{\rm Tr}_B\left[Q^+, e^{-\beta P^-(\alpha)}\right]Q^+=\alpha\beta
{\rm Tr}_B Q^+ Q^-e^{\beta (Q^-)^2}}
where, again, \fund\ has been used in the last step.
In terms of the eigenstates $M_i$ of equation \qpm\
(which we recall can be positive or negative) we find:
\eqn\sumb{Z(\beta)=\alpha\beta\sum_{i\in B}M_i\exp(-\beta M_i^2)+O(\alpha^2)}
where the sum over $i$ runs only over bosonic bound states.
Clearly, if all (or most) of the $M_i$ had the same sign, $Z(\beta\to 0)$
would diverge as $\exp(c/\beta)$. A finite limit would imply almost complete
cancellations between the different terms in the sum, and would
require large numbers of positive and negative $M_i$ at high $|M|$.

In the next section we shall study the signs of $M_i$ numerically.
We shall also numerically verify \ssbd. This is necessary because
in the derivation above we have used certain properties of the
Hilbert space
$\ch_\alpha=\{|B\rangle_\alpha, |F\rangle_\alpha\}$
which are not a priori guaranteed.
In particular, it is not obvious that operators like $H_0, Q^+Q^-$
act well on $\ch_\alpha$; in the `t Hooft model
\ref\thone{G. 't Hooft, \np {\bf B75} (1974) 461.}\
which is analogous to adjoint
QCD in some respects, similar operators are actually singular
in certain regions of parameter space. While we do not expect this
to be the case here, it seems useful to check \ssbd, at least in
simple examples.

\newsec{Numerical Results}

The system of equations involved in solving \bbb\ form an infinite number of
multivariable integral equations which have
so far resisted all attempts at an exact solution.
It is however possible to reduce this problem to the tractable
problem of diagonalizing finite
matrices.  This is done by discretizing
light -- cone momentum; we will not describe
the details of the discretization, which appear in \kl,
\ref\bdk{G. Bhanot, K. Demeterfi and I. Klebanov,
\pr {\bf D48} (1993) 4980.}.

Following the convention of \kl\ we write $P^+P^-$ in the form
\eqn\pppm{P^+P^- = xH_0+H_1}
where $x=m^2/g^2N$, $H_0$ is as in \ho, and
$H_1=\int_0^\infty{dk\over{k^2}}J_{ab}(-k)J_{ba}(k)$.
Note that $x$ is related to $\alpha$ in \sss\ by $x=\alpha+1$,
i.e. $x=1$ corresponds to the supersymmetric point.
Below we shall
investigate some aspects of
the mass spectrum as a function of $x$ at large $N$.  We also look at the
spectrum of $Q^+Q^-$, in particular the distribution of the signs of $M$
in \qpm.

There are two main difficulties with obtaining reliable
numerical results for the spectrum of QCD$_2$ coupled to adjoint matter:

\noindent 1) The size of the matrices one needs to diagonalize
(even at large $N$) increases rapidly with the cutoff. This is
especially problematic when one is studying highly excited states,
whose wavefunctions are (generically) rapidly varying with
momentum, or contain many quarks (or both).

\noindent 2) It is clear that the condition of normalizability
of the light -- cone wavefunction should play a crucial role
in selecting physical states and making the spectrum discrete.
In large $N$ adjoint QCD$_2$ there is much more room than in the `t Hooft model
\thone\ for states with finite norm at finite cutoff
(of course all states have finite norm then) to become
non -- normalizable as the cutoff is removed.
This is due to the presence of sectors with arbitrarily
many adjoint quarks. It is very difficult
in practice
to follow the states while increasing the cutoff and check whether
they survive in the continuum limit.

Nevertheless, it was shown in \kl, \bdk\ that for a few low
lying states these effects are numerically small. In particular,
the lowest lying excitation in the fermionic sector
contains to a high precision three adjoint quarks, while the bosonic
one has significant components only in the 2,4,6 quark sectors.
Thus, to verify \ssbd\ we diagonalized \pppm\ truncating to the above
mentioned sectors and continuing the results for the lowest
lying state from finite cutoff
($K=24$ for bosons and $K=25$ for fermions in the notation
of \bdk) to infinite one using a certain Pade
approximation. The results are exhibited
in Figs 1,2. The uncertainty in the masses
squared due to the various truncations
and extrapolations is estimated to be $2 - 3\%$.

In Fig. 1 we plot the masses squared of the lowest lying excitations
in the bosonic and fermionic sectors as a function of the ratio $x=m^2/g^2N$
in the region $0<x<1.5$. The mass of the fermionic bound state at zero
constituent
quark mass is calculated to be $5.72$. That of the boson is $10.77$.
For $x=1$ (the supersymmetric point) we obtain values of $25.73$ and $25.82$
respectively for the bosonic and fermionic states. All of the above
are in good agreement with \bdk. We see that to a good approximation
the numerical calculation reproduces the qualitative features of the theory
at least for the lowest lying state. Note also that as expected the individual
masses show non -- linear behavior away from the supersymmetric point.

In Fig. 2 we plot $M_F^2(x)-M_B^2(x)$. According to \ssbd\ this plot
should be a straight line near $x=1$; surprisingly, we find a straight
line over the full range of the graph. The deviation of the points from
 the straight
line fit of Fig. 2 is significantly less than the uncertainties mentioned
above. The slope of the straight line of Fig. 2 is $5.14$, whereas
\ssbd\ predicts \bdk: $\sqrt{25.8}=5.08$. The two agree to within the
accuracy of our numerical analysis.

The sign of the slope can also be determined by diagonalizing
$Q^+Q^-$. We find numerically that $Q^+Q^-$ is positive on the bosonic
lowest lying state, which means (using \ssbd) that the fermion is heavier
than the boson for $m^2>g^2N$ and vice versa. This again agrees
with known results (see Fig. 1). Note also that the surprising linearity
of $M_F^2-M_B^2$ extends all the way to infinite mass $(x\to\infty)$,
since we know that as $x\to\infty$, $M_B^2(x)\simeq 4x$,
$M_F^2(x)\simeq 9x$ (most of the meson mass
is then due to the masses of the constituents),
such that $M_F^2-M_B^2\simeq 5x$. Within the $2-3\%$
errors of our calculations $M_F^2-M_B^2$ is linear in this case for all $m$
between $0$ and $\infty$!

A second numerical check we have performed is related to the
evaluation of \sumb. We have looked at the distribution
of the signs of the eigenvalues of $Q^+Q^-$ to see whether
the rather drastic cancellations required by \ks\ are at all possible.
In Fig. 3 we show the absolute values of the eigenvalue spectrum of
$Q^+Q^-$ for a cutoff of $K=17$.  There are a total of 210 states in the
fermionic sector at this cutoff.  The graph shows that there are significant
numbers of states with both positive and negative eigenvalues.
To use this to calculate \sumb\ in the limit $\beta\to 0$
numerically would require a precise knowledge of the eigenvalues
$M_i$ and thus would involve cutoffs much higher than have been used
so far \bdk; the main lesson from fig. 3 is that it is quite
conceivable that the required cancellations may take place.

\newsec{The Landau -- Ginzburg description of QCD$_2$.}

In the previous sections we have derived rather mysterious
relations for the spectrum of the theory \a. The purpose of this section
is to present these results in a somewhat different light, which perhaps
will help explain their origin, as well as suggest ways of reaching a deeper
qualitative understanding of the theory.

It is well known that two dimensional QCD can be generalized to a theory with
an infinite number of couplings. This is easiest to see by replacing
$\cl={1\over 4g^2}F^2$ by
$\cl=\phi F-g^2\phi^2$, where $F={1\over2}\epsilon^{\mu\nu}F_{\mu\nu}$
and $\phi$ is an auxiliary scalar field in the adjoint representation
of the gauge group. In pure QCD this is natural since the theory
with $g=0$ then becomes a well known topological field theory. Of course,
one can now generalize to
$\cl=\phi F-f(\phi)$, with $f$ any function of $\phi$. In the presence
of adjoint fermions one may also add terms like $f_1(\phi)\bar\psi f_2(\phi)
\psi$. In general, all these couplings will break supersymmetry \ae\
explicitly, however one can preserve supersymmetry in a rather natural
way. In fact, given any superpotential:
\eqn\wphi{W(\Phi)=\sum_n{a_n\over n+1}\Phi^{n+1}}
we shall find that that the Lagrangian:
\eqn\supl{\cl=\phi F-\left[W^\prime(\phi)\right]^2+\bar\psi\gamma^\mu D_\mu
\psi+\sqrt{N}\bar\psi\{\psi W^{\prime\prime}(\phi)\}}
is supersymmetric.
In \supl\ we have introduced the notation $\{\psi f(\phi)\}$ which is
defined by:
\eqn\defbracket{
\{\psi\phi^n\}\equiv{1\over n+1}\sum_{i=0}^n\phi^i\psi\phi^{n-i}.}

To prove this assertion one follows the steps of section 2,
picking $A_-=0$, such that :
\eqn\solv{\cl=\phi\partial_-A_+-\left[W^
\prime(\phi)\right]^2+i\psi\partial_+\psi
+i\bar\psi\partial_-\bar\psi-2\sqrt{N}i\bar\psi\{\psi W^{\prime\prime}(\phi)\}+
A_+J^+.}
Note that one can integrate over $A_+$; this sets
\eqn\setphi{\phi={1\over\partial_-}J^+.}
The theory of section 2 (with $m^2=g^2N$)
is described by the superpotential $W(\phi)={1\over2}g\phi^2$.
One can easily write the form of $P^-$ (after eliminating $\bar\psi$ as in
section 2):
\eqn\pminusnew{P^-_W=\int dx^-\left(-i N\{\psi W^{\prime\prime}(\phi)\}
{1\over\partial_-}
\{\psi W^{\prime\prime}(\phi)\}
+\left(W^\prime(\phi)\right)^2\right).}
Note that the supersymmetry transformation generated by $Q^+$, \ae\ (which
of course should be independent of the superpotential) take in terms of
$\phi/\sqrt N$
\setphi\ and $\psi$, the familiar form:
\eqn\newsus{\delta\left({\phi\over\sqrt N}\right)=-i\epsilon\psi;\;\;\;
\delta\psi=\epsilon\partial_-\left({\phi\over\sqrt N}\right).}
These are precisely the transformation laws of a scalar superfield,
and of course $P^-_W$ is essentially the light -- cone Hamiltonian
for that case. The main difference is that in our case $\phi$ is not an
independent fluctuating field; eqn. \setphi\ relates it to $\psi$.
At any rate, it easy to check that $[Q^+, P_W^-]=0$.
To do that one verifies that
\eqn\commut{\eqalign{[Q^+, W^\prime(\phi)]=&-i\sqrt{N}\{\psi
W^{\prime\prime}(\phi)\}\cr
\{Q^+,\{\psi W^{\prime\prime}(\phi)\}\}=&{1\over\sqrt N}
\partial_-(W^\prime(\phi)).\cr}}
We shall not describe the details here which are very similar to
the case of scalar superfields.
It is more difficult to construct $Q^-$ and check $(Q^-)^2=-P^-$
but a likely guess is:
\eqn\qminus{Q^-=\int dx^-\{\psi W^\prime(\phi)\}.}
It is easy, using \newsus, to check that $\{Q^+, Q^-\}=0$. Again,
\qminus\ has the same form as that for a scalar
superfield; it reduces to \qm\ for
a quadratic superpotential.

There are two main reasons why the generalized gauge theories \supl\ may be
useful to consider:

\noindent{}1) We have exhibited an analogy of the physics of our theory
to that of an interacting superfield. The latter is known to describe
supersymmetric minimal models and flows between them
\ref\zam{A. Zamolodchikov, Sov. J. Nucl. Phys. {\bf 44} (1986) 44.}
\ref\kms{D. Kastor, E. Martinec, S. Shenker, Nucl. Phys. {\bf B316}
(1989) 590.}; it is integrable and possesses an infinite number of
conserved currents. It would be interesting if using this analogy one could
learn more about possible extended symmetries in the gauge theories. In
particular,
the theory considered in sections 2 - 4 has a quadratic superpotential;
it is possible that the mass splitting relations \ssbd\
can be related to similar relations in the free theory,
and are a sign of a hidden free field structure in the gauge theory.

\noindent{}2) One of the main purposes of this work is to try and relate
QCD$_2$ coupled to adjoint matter to string theory. There is some
evidence \ku\ that aspects of the model sensitive to the physics of highly
excited
states (such as high temperature behavior) are indeed closely related
to string theory. However, to have a continuous worldsheet description
of the theory at all scales it seems necessary to fine tune couplings
such that even low lying eigenstates of the Hamiltonian consist of many
quarks. As mentioned above, this is not the case for a generic
superpotential -- pair production is dynamically suppressed \bdk.
By fine tuning $W(\Phi)$ \supl\ it is possible that critical points
can be found at which the average number of quarks in the wavefunctions
of all low lying states diverges\foot{Such critical points are known to
be relevant for continuum physics in the matrix model description of
$1+1$ dimensional string theory \gm, and one expects on general grounds
that existence of a critical point should be necessary for a
string description in any gauge theory
with propagating degrees of freedom in the adjoint representation.}.

In addition, one can study the rich structure of this theory as one varies
$W$; one expects spontaneous breaking of SUSY, non trivial
massless spectra, etc, in analogy to \kms.

\newsec{Conclusions}

In this paper we have continued the program of ref. \ku, and tried to
obtain further analytical information about QCD$_2$ coupled
to adjoint fermions. Our main result is the mass splitting formula,
eqn. \ssbd. It is rather remarkable to find in a complicated dynamical
system with an infinite number of bound states (as $N\to\infty$) such
simple relations between masses. These universal relations, which hold
for all $N$, resemble naive quark model predictions. Indeed,
if the bosonic bound states consisted of $2n$ quarks and had total
mass $M_B=2 n m+U_B$ with $m$ the constituent quark mass and $U_B$
the binding energy, and were paired by SUSY with fermionic states
consisting of $2n\pm1$ quarks with mass $M_F=(2n\pm1)m+U_F$, such that
at a certain $m=m_{\rm susy}$ the two were degenerate $M_B=M_F$, one would
indeed have for $m=m_0+\delta m$, $M_B-M_F=\pm \delta m$.
Of course, this picture is naive in many respects; it is non --
relativistic, ignores pair production, etc. Nevertheless, the universality
of the mass splittings \ssbd\ seems to suggest some hidden
simplicity in the model.
Perhaps the analogy pointed out in section 5 between QCD and field theory
of a free massive superfield can be used to understand the origin of these
results.
The algebraic structure, in
particular the role of the operators $Q_n$ \qn\ and possible higher
conserved currents (see section 5) certainly
requires better understanding.

We have also studied the quantity $Z(\beta)$ \zb; although
the results are inconclusive, numerically it seems that
certain string constraints which would hold in any conventional
string theory are at least not ruled out. If it is found that
these constraints do hold, this would be a strong indication
of a stringy structure of the theory.

Despite this progress, the most important problems remain unsolved.
The main problem is to find the spectrum of the theory (at least
at large $N$). The main difficulty is that most eigenstates
one may write down are non -- normalizable (due to the presence
even at large $N$ of arbitrarily high quark number sectors in the
light -- cone wavefunction), and the choice of normalizable
eigenstates requires detailed knowledge of the light -- cone wavefunctions;
this is clearly not the way to proceed. This problem makes the
numerical analysis difficult as well.

Of course, it would be interesting
to find a string description of this model without solving it, and
there are some indications that one exists \ku. In particular,
it seems promising to look at the many adjoint quark components
of light -- cone wavefunctions; in this situation the quarks effectively
form a string with a continuous distribution of light -- cone momentum.
A promising idea is to look for critical points at which the average number
of quarks in a hadron diverges, by fine tuning the superpotential $W$
of section 5. Supersymmetry will insure that no tachyons appear.
All these, and other issues must be left for future studies.

\bigbreak\bigskip\centerline{{\bf Acknowledgements}}\nobreak

D. K. thanks Y. Frishman and E. Rabinovici for interesting comments
and the Weizmann Institute for hospitality.
This work was partially supported by a DOE OJI grant and
the L. Block foundation.  J.B. is supported by a NSF Graduate Fellowship.

\listrefs

\bigbreak

\centerline{FIGURE CAPTIONS}
\bigskip

\item
{\bf Fig.1.}The masses squared of the lowest lying fermionic and bosonic
states as a function of $x=m^2/g^2N$.
\item
{\bf Fig.2.}The difference in the masses squared of the lowest lying fermionic
and bosonic states as a function of $x$.
\item
{\bf Fig.3.}The distribution of eigenvalues of $Q^+Q^-$ for a cutoff of 17.

\bye